\documentclass[journal,10pt]{IEEEtran}
\makeatletter
\def\endthebibliography{%
    \def\@noitemerr{\@latex@warning{Empty `thebibliography' environment}}%
    \endlist
}
\makeatother

\usepackage{cite}

\usepackage[pdftex]{graphicx}
\usepackage[caption=false,font=footnotesize]{subfig}

\usepackage{amsmath}
\usepackage{mathtools, cuted}
\usepackage{amssymb}
\usepackage{bm}
\usepackage{mathrsfs}
\usepackage{breqn}

\usepackage{pifont}
\usepackage{xcolor}
\usepackage{url}
\usepackage{lettrine}
\usepackage{lipsum}
\usepackage{siunitx}
\usepackage{soul}
\usepackage{array}
\usepackage[inline]{enumitem}
\usepackage{epsfig}
\usepackage{filecontents}

\usepackage{algpseudocode}
\usepackage{algorithm, tabularx}
\usepackage{multirow}
\newcolumntype{L}[1]{>{\raggedright\let\newline\\\arraybackslash\hspace{0pt}}m{#1}}
\newcolumntype{C}[1]{>{\centering\let\newline\\\arraybackslash\hspace{0pt}}m{#1}}
\newcolumntype{R}[1]{>{\raggedleft\let\newline\\\arraybackslash\hspace{0pt}}m{#1}}
\newlength{\maxwidth}

\makeatletter
\newcommand{\multiline}[1]{%
	\begin{tabularx}{\dimexpr\linewidth-\ALG@thistlm}[t]{@{}X@{}}
		#1
	\end{tabularx}
}
\makeatother
\algdef{SE}[SUBALG]{Indent}{EndIndent}{}{{\algorithmicend\ }}
\algtext*{Indent}
\algtext*{EndIndent}

\usepackage{amsthm}

\theoremstyle{remark}

\begin{document}

\title{\LARGE Robust Joint Beamforming and Configuration Design in\\FARIS-Aided Systems}

\author{Hong-Bae Jeon,~\IEEEmembership{Member,~IEEE}, Yonghwi Kim,~\IEEEmembership{Member,~IEEE,} Hyung-Joo Moon,~\IEEEmembership{Member,~IEEE,} and\\Kai-Kit Wong,~\IEEEmembership{Fellow,~IEEE} 
\thanks{\textit{(Corresponding Author: Hyung-Joo Moon)}}%
\thanks{H.-B. Jeon is with the School of Electronic Engineering, Soongsil University, Seoul, Korea (e-mail: hongbae08@ssu.ac.kr).}%
\thanks{Y. Kim is with the Department of Electronics and Electrical Engineering, Dankook University, Yongin, Korea (e-mail: eric\_kim@dankook.ac.kr).}
\thanks{H.-J. Moon is with the School of Integrated Technology, Yonsei University, Seoul, Korea (e-mail: moonhj@yonsei.ac.kr).}%
	\thanks{K.-K. Wong is with the Department of Electronic and Electrical Engineering, University College London, WC1E 6BT London, U.K., and also with the Yonsei Frontier Laboratory, Yonsei University, Seoul, Korea (e-mail: kai-kit.wong@ucl.ac.uk).}
}

\maketitle

\begin{abstract}
In this paper, we propose a robust transmission design for multi-user systems assisted by a fluid active reconfigurable intelligent surface (FARIS), which enables both active reflection and dynamic port selection and thereby offers enhanced flexibility, under imperfect channel state information (CSI). We formulate a robust minimum sum-rate maximization problem by jointly optimizing the base station beamformer, the utilized FARIS coefficients, and the active element selection, while explicitly accounting for CSI errors and practical power constraints. The resulting problem is inherently nonconvex due to the coupled optimization variables and discrete port-selection structure. To tackle this challenge, we first reformulate the original problem via a weighted minimum mean square error (WMMSE) approach and then devise an alternating optimization (AO) framework, where each resulting subproblem admits efficient solutions and the overall algorithm converges to a stationary point. Simulation results demonstrate that the proposed robust FARIS scheme consistently outperforms conventional designs, highlighting the effectiveness of jointly leveraging degree-of-freedom (DoF) enhancement and active signal amplification under CSI uncertainty.
\end{abstract}

\begin{IEEEkeywords}
Fluid active reconfigurable intelligent surface (FARIS), channel state information (CSI) error, robust beamforming.
\end{IEEEkeywords}

\IEEEpeerreviewmaketitle

\section{Introduction}
\label{sec:intro}
Reconfigurable intelligent surface (RIS) have recently emerged as a promising technology for enhancing spectral efficiency and coverage in next-generation wireless networks by enabling programmable manipulation of the radio propagation environment~\cite{risspm}. By appropriately configuring the phase responses of a large number of low-cost reflecting elements, RIS-assisted systems can reshape signal propagation paths without requiring additional radio-frequency chains at the surface~\cite{TMH, nfris}. However, conventional RIS architectures are inherently passive, and their performance gains are fundamentally limited by severe double-fading attenuation~\cite{DF}, especially in scenarios with unfavorable propagation conditions~\cite{HBRIS}.

To overcome these limitations, active-RIS (ARIS) have been proposed~\cite{aris1, aris5}, in which each reflecting element is equipped with an active amplifier. By enabling both phase control and signal amplification, ARIS can significantly enhance the received signal strength and mitigate the double-fading effect. Motivated by these advantages, recent studies have investigated joint beamforming and reflection design for ARIS-assisted systems~\cite{aris5, aris8}, demonstrating notable performance improvements over passive-RIS counterparts. Nevertheless, the amplification capability inevitably introduces additional thermal noise and imposes stringent radiated power constraints at ARIS, which complicate the system design, particularly under imperfect channel state information (CSI)~\cite{aris7, aris8}.

In parallel, fluid-RIS (FRIS) have been introduced to further expand the design space of RIS-assisted systems~\cite{FRISlook, FRISmag}, motivated by fluid antenna systems~\cite{fas}. Unlike conventional RIS with fixed element locations, FRIS dynamically activates only a subset of ports from a dense grid of candidate locations, thereby exploiting location diversity in addition to phase control. This fluid structure enables additional spatial degree-of-freedom (DoF) and improved adaptability to channel variations~\cite{FRISonoff}. However, FRIS architectures remain passive, and their performance is still constrained in scenarios where severe path loss or weak signal conditions dominate.

More recently, the concept of a fluid-active-RIS (FARIS), first introduced by \textit{\textbf{Jeon}}~\cite{FARIS}, has been proposed to unify the advantages of active signal amplification and fluid port selection within a single architectural framework. By jointly enabling signal amplification and dynamic port activation, FARIS provides a substantially richer design space than conventional architectures. This hybrid structure allows the system to not only enhance favorable signal paths through amplification but also selectively exploit advantageous spatial locations, offering a powerful means to improve performance in challenging propagation environments. Despite its considerable promise, the practical design of reconfigurable-surface-based systems, including FARIS, faces inherent challenges in obtaining accurate CSI, primarily due to the absence of baseband processing at the surface and the presence of a large number of candidate ports~\cite{CE3, nearris}.

Therefore, this paper investigates a robust transmission design for multi-user systems assisted by FARIS under imperfect CSI. In particular, we aim to fully exploit the joint benefits of active amplification and enhanced spatial DoF, while explicitly accounting for channel uncertainty and practical hardware constraints. The main contributions are summarized as follows:
\begin{itemize}
\item We adopt a statistical CSI error model to explicitly characterize the impact of channel uncertainty on both the direct and FARIS-assisted links. By considering the worst-case effect of CSI errors, we formulate a robust minimum sum-rate maximization problem that jointly optimizes the beamformer, the utilized FARIS coefficients, and the active port selection, while satisfying practical power constraints at both the base station and the FARIS.
\item The resulting problem is inherently nonconvex due to the coupling between continuous beamforming variables, active surface coefficients, and the sparsity-inducing port selection structure. To address this challenge, we develop a weighted minimum mean square error (WMMSE)-based reformulation, followed by an alternating optimization (AO) framework, which decomposes the original nonconvex problem into a sequence of convex or efficiently solvable subproblems.
\item Numerical results demonstrate that the proposed robust FARIS design outperforms the benchmarks, highlighting the effectiveness of jointly exploiting active amplification and additional spatial DoF under CSI uncertainty.
\end{itemize}

\begin{figure}[t]
  \begin{center}
    \includegraphics[width=0.95\columnwidth,keepaspectratio]{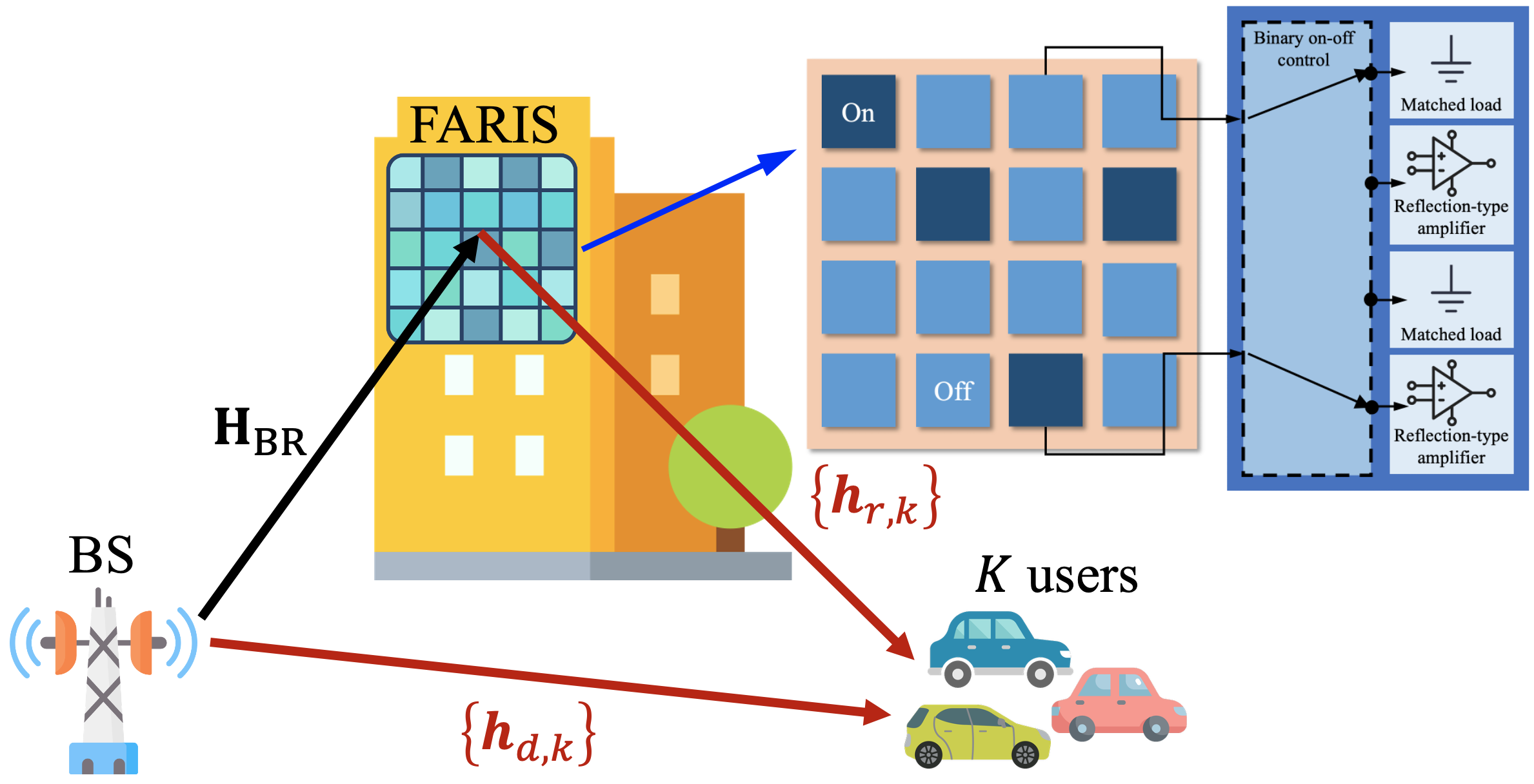}
    \caption{FARIS-aided multi-user downlink system with the schematic illustration of FARIS.}
    \label{fig_sys}
  \end{center}
\end{figure}
\section{System Model}
\label{sec:sys}
As illustrated in Fig.~\ref{fig_sys}, we consider a downlink multi-user system assisted by a FARIS, composed of $M=M_x\times M_x$ reflective elements uniformly distributed over $W_x\lambda \times W_x\lambda$-size square surface. Herein, $\lambda$ denotes the carrier wavelength and $W_x$ represents the $\lambda$-normalized aperture size. The resulting inter-element spacing is given by $d=\frac{W_x\lambda}{M_x}$. Due to the finite spacing between elements, spatial correlation is captured by a correlation matrix $\mathbf J\in\mathbb R^{M\times M}$, whose $ij$th entry is modeled using Jakes' correlation model as $J_{ij}=j_0\left(\frac{2\pi d_{ij}}{\lambda}\right)$~\cite{FRISonoff, FRISpa}, where $j_0(\cdot)$ denotes the zero-order spherical Bessel function of the first kind, and $d_{ij}$ represents the physical distance between the $i$ and $j$th elements. Following~\cite{FRISonoff, FRISmag}\footnote{Although the referenced model is developed for FRIS, it is adopted to FARIS~\cite{FARIS} since it shares the same underlying hardware architecture, with the only difference being the additional amplification module.} and illustrated in Fig.~\ref{fig_sys}, each FARIS element is interpreted as a port of a fluid antenna structure, operates in one of two modes; In \textbf{on} state, it actively interacts with the incident electromagnetic wave and applies controllable amplitude and phase modifications. In \textbf{off} state, it is terminated with a matched load, effectively isolating it from the impinging signal and preventing any reflection.

The base station (BS) is equipped with $N$ antennas and serves $K$ single-antenna users. Among $M$ candidate ports of FARIS, only $M_o$ ports are activated at any given time. Under the conventional discrete port-selection model, this activation is described by a selection matrix $\mathbf S^{\mathrm T}\in\{0,1\}^{M\times M_o}$:
\begin{equation}
\label{eq:Sm}
\mathbf S^{\mathrm T}
\triangleq
\big[\mathbf e_{i_1} \cdots \mathbf e_{i_{M_o}}\big]~(\forall i_m\in\{1,\cdots,M\}),
\end{equation}
where $\mathbf e_i\in\mathbb R^M$ denotes the canonical basis vector; the $i$th component is 1 and others are zero. This construction ensures that only the ports indexed by $\{i_m\}_{m=1}^{M_o}$ are on, while the requirement that all columns of $\mathbf S^{\mathrm T}$ be distinct guarantees unique port activation without redundancy.

For the selected $M_o$ ports, signal manipulation is realized through per-port phase shifts $\{\phi_i\in[0,2\pi)\}$ and amplification gains $\{g_i\in[0,g_{\max}]\}$. Specifically, let $\boldsymbol\Phi=\mathrm{diag}\left(\{e^{j\phi_i}\}_{i=1}^{M_o}\right)$ and $\mathbf G=\mathrm{diag}\left(\{g_1\}_{i=1}^{M_o}\right)$, and define the coefficient vector $\mathbf v=[v_1 \cdots v_{M_o}]^{\mathrm{T}}\in\mathbb C^{M_o}$ such that $\mathrm{diag}(\mathbf v)=\boldsymbol\Phi\mathbf G$. The resulting FARIS operator can then be $\mathbf A_{\mathrm F}=\mathbf J^{1/2}\mathbf S^{\mathrm T}\mathrm{diag}(\mathbf v)\mathbf S\mathbf J^{1/2}$.

To obtain a more compact and selection-free representation, we define an equivalent per-port coefficient vector $\mathbf w\in\mathbb C^{M}$ as $\mathbf w\triangleq \mathbf S^{\mathrm T}\mathbf v$, or equivalently,
\begin{equation}
\label{eq:w_}
w_m=
\begin{cases}
v_n & (\exists n\in\{1, \cdots, M_o\} ~\text{s.t.}~ m=i_n),\\
0 & (\text{otherwise}),
\end{cases} (\forall m).
\end{equation}
By construction, $\mathbf w$ is $M_o$-sparse, i.e., $\|\mathbf w\|_0=M_o$. Using this definition, we can rewrite $\mathbf{A}_{\mathrm{F}}$ as
\begin{equation}
\label{eq:AF_w_derivation}
\begin{aligned}
\mathbf A_{\mathrm F}&=\mathbf J^{1/2}\mathbf S^{\mathrm T}\mathrm{diag}(\mathbf v)\mathbf S\mathbf J^{1/2}\\
&\overset{(a)}{=}\mathbf J^{1/2}\mathrm{diag}(\mathbf S^{\mathrm T}\mathbf v)\mathbf J^{1/2}=\mathbf J^{1/2}\mathrm{diag}(\mathbf w)\mathbf J^{1/2},
\end{aligned}
\end{equation}
where $(a)$ follows from $\mathbf S^{\mathrm T}\mathrm{diag}(\mathbf v)\mathbf S=\sum_{m=1}^{M_o} v_m\mathbf e_{i_m}\mathbf e_{i_m}^{\mathrm T} =\mathrm{diag}(\mathbf S^{\mathrm T}\mathbf v)$.

To avoid combinatorial optimization over $\mathbf S$, we relax the binary port activation by introducing $\mathbf p\in[0,1]^M$ with a fixed activation budget $\mathbf 1^{\mathrm T}\mathbf p = M_o$. We couple $\mathbf p$ with $\mathbf w$ via the perspective constraint: $|w_m|^2 \le g_{\max}^2 p_m~(m=1,\cdots,M)$, where $g_{\max}$ is the maximum amplification gain. When $\mathbf p\in\{0,1\}^M$, it enforces $w_m=0$ if $p_m=0$ and $|w_m|\le g_{\max}$ if $p_m=1$, thereby recovering the discrete port selection.

Let $\mathbf H_{\mathrm{BR}}\in\mathbb C^{M\times N}$ denote the BS-FARIS channel. The direct BS-user channel $\mathbf h_{d,k}$ and FARIS-user channel $\mathbf h_{r,k}$ for user $k$ are modeled by the statistical CSI error model~\cite{mfris}:
\begin{equation}
\label{eq:hd_model}
\mathbf h_{j,k} = \hat{\mathbf h}_{j,k} + \Delta \mathbf h_{j,k},
~\Delta\mathbf h_{j,k}\sim\mathcal{CN}(\mathbf 0,\sigma_{j,k}^2\mathbf I)~(j\in\{d,r\}),
\end{equation}
where $\hat{\mathbf h}_{d,k}$ and $\hat{\mathbf h}_{r,k}$ are the channel estimates. Let $\mathbf F=[\mathbf f_1 \cdots \mathbf f_K]\in\mathbb C^{N\times K}$ be the BS beamformer with $\|\mathbf F\|_F^2\le P_B$. With unit-power symbols $\mathbf s=[s_1 \cdots  s_K]^{\mathrm{T}}\in\mathbb C^K$ and mutual independence, the received signal at user $k$ is
\begin{equation}
\label{eq:yk}
y_k
=
\sum_{i=1}^{K}
\Big(
\mathbf h_{d,k}^{*}
+
\mathbf h_{r,k}^{*}\mathbf A_{\mathrm F}(\mathbf w)\mathbf H_{\mathrm{BR}}
\Big)\mathbf f_i s_i
+
\mathbf h_{r,k}^{*}\mathbf A_{\mathrm F}(\mathbf w)\mathbf z
+
n_k,
\end{equation}
where $\mathbf z\sim\mathcal{CN}(\mathbf 0,\sigma_r^2\mathbf I)$ is the FARIS thermal noise and $n_k\sim\mathcal{CN}(0,\sigma_k^2)$ is the receiver noise. Thereby, the instantaneous FARIS output power is
\begin{equation}
\label{eq:PF}
\underbrace{\big\|\mathbf A_{\mathrm F}(\mathbf w)\mathbf H_{\mathrm{BR}}\mathbf F\big\|_F^2
+
\sigma_r^2\operatorname{tr}\Big(\mathbf A_{\mathrm F}(\mathbf w)\mathbf A_{\mathrm F}(\mathbf w)^{*}\Big)}_{\triangleq P_{\mathrm F}(\mathbf F,\mathbf w)}
\le P_{\max}.
\end{equation}
In addition to $P_{\mathrm F}(\mathbf F,\mathbf w)$, the operation of FARIS also incurs hardware power consumption associated with the selected fluid-active ports. Specifically, based on the hardware architecture and circuit model illustrated in Fig.~\ref{fig_sys}~\cite{FARIS}, the circuit power consumption of FARIS consists of two components:
i) $P_c$, representing the logical control and switching power consumed by each candidate element, and
ii) $P_{\mathrm{DC}}$, denoting the direct current (DC) bias power required for active reflection~\cite{aris5}.
Since the logical control circuitry is required for all $M$ candidate elements, whereas the active reflection branch is given only for the selected $M_o$ FARIS ports, the total FARIS power consumption can be expressed as
\begin{equation}
\label{eq:Pfaris_total}
P_{\mathrm{FARIS}}^{\mathrm{tot}}
\triangleq
MP_c + M_o P_{\mathrm{DC}} + \xi P_{\mathrm F}(\mathbf F,\mathbf w),\footnote{Note that when $M_o=M$, the model reduces to the conventional ARIS power consumption model in~\cite{aris5}.}
\end{equation}
where $\xi \triangleq \frac{1}{\upsilon}$ and $\upsilon\in(0,1]$ denotes the amplifier efficiency. Accordingly, under the total FARIS power budget $P_{\max,t}$, the reflection design must satisfy $P_{\mathrm{FARIS}}^{\mathrm{tot}} \le P_{\max,t}$, equivalently:
\begin{equation}
\label{eq:Pris_constraint}
P_{\mathrm F}(\mathbf F,\mathbf w)
\le
\upsilon\big(P_{\max,t}-MP_c-M_oP_{\mathrm{DC}}\big)
\triangleq
P_{\max}.
\end{equation}
Now define the effective estimated composite channel $\hat{\mathbf h}_k^{*}(\mathbf w)\triangleq \hat{\mathbf h}_{d,k}^{*}+\hat{\mathbf h}_{r,k}^{*}\mathbf A_{\mathrm F}(\mathbf w)\mathbf H_{\mathrm{BR}}$. Under the statistical CSI error, the minimum achievable rate $R_{\min,k}$ of user $k$ becomes
\begin{equation}
\label{eq:Rmin_new}
R_{\min,k}=\log_2\left(1+\frac{\big|\hat{\mathbf h}_k^{*}(\mathbf w)\mathbf f_k\big|^2}{\Gamma_k(\mathbf F,\mathbf w)}\right),
\end{equation}
where $\Gamma_k(\mathbf F,\mathbf w)=\sum_{i\neq k}\big|\hat{\mathbf h}_k^{*}(\mathbf w)\mathbf f_i\big|^2+\sum_{i=1}^{K}\Big(\sigma_{d,k}^2\|\mathbf f_i\|_2^2+\sigma_{r,k}^2\|\mathbf A_{\mathrm F}(\mathbf w)\mathbf H_{\mathrm{BR}}\mathbf f_i\|_2^2\Big) +\sigma_r^2\big\|\hat{\mathbf h}_{r,k}^{*}\mathbf A_{\mathrm F}(\mathbf w)\big\|_2^2+\sigma_r^2\sigma_{r,k}^2\big\|\mathbf A_{\mathrm F}(\mathbf w)\big\|_F^2+\sigma_k^2$ is the interference-plus-noise term~\cite[Appendix A]{aris8}. The robust FARIS-aided transmission design is therefore formulated as
\begin{equation}
\label{prob:robust_relax}
\begin{aligned}
&\max_{\mathbf F,\mathbf w,\mathbf p}~ \sum_{k=1}^{K} R_{\min,k} \\
\mathrm{s.t.}~
& \|\mathbf F\|_F^2 \le P_B,~ P_{\mathrm F}(\mathbf F,\mathbf w)\le P_{\max},\\
& |w_m|^2 \le g_{\max}^2 p_m ~( 0\le p_m\le 1, \forall m),~\mathbf 1^{\mathrm T}\mathbf p=M_o.
\end{aligned}
\end{equation}
Problem~\eqref{prob:robust_relax} remains nonconvex due to the coupled beamforming and utilized FARIS coefficients. We address this challenge via developing an AO framework.
\section{Proposed Framework}
\subsection{WMMSE reformulation with auxiliary variables}
\label{subsec:wmmse}
Define the mean-squared-error (MSE) of user $k$ under a linear receiver $u_k\in\mathbb C$ as $e_k(u_k,\mathbf F,\mathbf w)
\triangleq\mathbb E\left[\big|s_k-u_k^{*}y_k\big|^2\right]$. Substituting~\eqref{eq:yk} and together with the error statistics yields the following closed form~\cite{aris8}:
\begin{equation}
\label{eq:MSE_closed_new}
\begin{aligned}
e_k=&\Big|1-u_k^{*}\hat{\mathbf h}_{k}^{*}(\mathbf w)\mathbf f_k\Big|^2
+\sum_{i\neq k}\Big|u_k^{*}\hat{\mathbf h}_{k}^{*}(\mathbf w)\mathbf f_i\Big|^2 \\
&+\sum_{i=1}^{K}
\Big(
\sigma_{d,k}^2|u_k|^2\|\mathbf f_i\|_2^2
+\sigma_{r,k}^2|u_k|^2\|\mathbf A_{\mathrm F}(\mathbf w)\mathbf H_{\mathrm{BR}}\mathbf f_i\|_2^2
\Big)\\
&+\sigma_r^2|u_k|^2\big\|\hat{\mathbf h}_{r,k}^{*}\mathbf A_{\mathrm F}(\mathbf w)\big\|_2^2
+\sigma_r^2\sigma_{r,k}^2|u_k|^2\big\|\mathbf A_{\mathrm F}(\mathbf w)\big\|_F^2\\
&+\sigma_k^2|u_k|^2 .
\end{aligned}
\end{equation}
For any $\nu_k\in\mathbb R_{++}$ and $u_k\in\mathbb C$, the following bound of~\eqref{eq:Rmin_new} holds by the concavity of the right-hand side (RHS) with respect to $(\nu_k, u_k)$~\cite[Proposition 1]{aris8}:
\begin{equation}
\label{eq:wmmse_lb_new}
\log_2\left(1+\frac{|\hat{\mathbf h}_{k}^{*}(\mathbf w)\mathbf f_k|^2}{\Gamma_k(\mathbf F,\mathbf w)}\right)
\ge\underbrace{ \log_2 \nu_k-\nu_k e_k(u_k,\mathbf F,\mathbf w)+c}_{\triangleq f(u_k , \nu_k)},
\end{equation}
where $c=\frac{1}{\ln 2}-\log_2\left(\frac{1}{\ln 2}\right)$. Note that unlike the conventional WMMSE formulation~\cite{wmmse}, it explicitly accounts for FARIS-related CSI errors, and establishes a relationship between $R_{\min,k}$ and the corresponding weighted MSE for each user.

For fixed $(\mathbf F,\mathbf w)$, maximizing concave $f(u_k, \nu_k )$ by solving $\frac{\partial f}{\partial u_k}=0$ and $\frac{\partial f}{\partial \nu_k}=0$ respectively yields:
\begin{equation}
\label{eq:u_nu_update_new}
u_k^\star=\frac{\hat{\mathbf h}_{k}^{*}(\mathbf w)\mathbf f_k}{|\hat{\mathbf h}_{k}^{*}(\mathbf w)\mathbf f_k|^2+\Gamma_k(\mathbf F,\mathbf w)}, \nu_k^\star=\frac{1}{\ln 2}\Big(1+\frac{|\hat{\mathbf h}_{k}^{*}(\mathbf w)\mathbf f_k|^2}{\Gamma_k}\Big),
\end{equation}
and by substituting into $f(u_k, \nu_k)$, we get $f(u_k^\star, \nu_k^\star)=R_{\min, k}$, which is the left-hand side (LHS) of~\eqref{eq:wmmse_lb_new}. 
Hence, we consider:
\begin{equation}
\label{prob:WMMSE_relax}
\begin{aligned}
&\max_{\boldsymbol{\nu},\mathbf u,\mathbf F,\mathbf w,\mathbf p}~ \sum_{k=1}^{K}(\log_2 \nu_k-\nu_k e_k(u_k,\mathbf F,\mathbf w))\\
\mathrm{s.t.}~
& \|\mathbf F\|_F^2 \le P_B,P_{\mathrm F}(\mathbf F,\mathbf w)\le P_{\max},\\
& |w_m|^2 \le g_{\max}^2 p_m, ( 0\le p_m\le 1, \forall m),\mathbf 1^{\mathrm T}\mathbf p=M_o.
\end{aligned}
\end{equation}
From~\eqref{prob:WMMSE_relax}, we propose the following AO framework:
\subsection{AO updates}
\subsubsection{Update of $(\boldsymbol{\nu},\mathbf u)$}
Each pair $(u_k,\nu_k)$ is updated by~\eqref{eq:u_nu_update_new}.
\subsubsection{Update of $\mathbf F$}
Fixing other variables,~\eqref{prob:WMMSE_relax} reduces to a convex quadratically constrained quadratic program (QCQP):
\begin{equation}
\label{prob:F_sub_new}
\min_{\mathbf F}~
\sum_{k=1}^{K}\nu_k e_k(u_k,\mathbf F,\mathbf w)~\mathrm{s.t.}~
\|\mathbf F\|_F^2\le P_B, P_{\mathrm F}(\mathbf F,\mathbf w)\le P_{\max},
\end{equation}
which can be efficiently solved using CVX solvers~\cite{boyd}.
\subsubsection{Update of $(\mathbf w,\mathbf p)$}
Fix $(\boldsymbol{\nu},\mathbf u,\mathbf F)$, define $\mathbf x_i \triangleq \mathbf J^{1/2}\mathbf H_{\mathrm{BR}}\mathbf f_i\in\mathbb C^{M},~\mathbf a_k \triangleq \mathbf J^{1/2}\hat{\mathbf h}_{r,k}\in\mathbb C^{M}$, and introduce the lifted variable $\mathbf W\triangleq \mathbf w\mathbf w^{*}\succeq \mathbf 0$. Using $\mathbf x^{*}\mathrm{diag}(\mathbf w)^{*}\mathbf J\mathrm{diag}(\mathbf w)\mathbf x=\operatorname{tr}\Big(\big(\mathbf J\odot(\mathbf x\mathbf x^{*})^{\mathrm T}\big)\mathbf W\Big)$, we obtain
\begin{equation}
\label{eq:trace_full_terms_corrected}
\begin{aligned}
\big\|\mathbf A_{\mathrm F}(\mathbf w)\mathbf H_{\mathrm{BR}}\mathbf f_i\big\|_2^2
&=\operatorname{tr}\big(\tilde{\mathbf Q}_i\mathbf W\big),
\tilde{\mathbf Q}_i\triangleq \mathbf J\odot(\mathbf x_i\mathbf x_i^{*})^{\mathrm T}\in\mathbb {C}^{M\times M},\\
\big\|\hat{\mathbf h}_{r,k}^{*}\mathbf A_{\mathrm F}(\mathbf w)\big\|_2^2
&=\operatorname{tr}\big(\tilde{\mathbf R}_k\mathbf W\big),
\tilde{\mathbf R}_k\triangleq \mathbf J\odot(\mathbf a_k\mathbf a_k^{*})\in\mathbb {C}^{M\times M},\\
\big\|\mathbf A_{\mathrm F}(\mathbf w)\big\|_F^2
&=\operatorname{tr}\big(\tilde{\mathbf T}\mathbf W\big),
\tilde{\mathbf T}\triangleq \mathbf J\odot\mathbf J\in\mathbb {C}^{M\times M}.
\end{aligned}
\end{equation}

For each $(k,i)$, define $b_{k,i}\triangleq \hat{\mathbf h}_{d,k}^{*}\mathbf f_i,~\mathbf d_{k,i}\triangleq \bar{\mathbf a}_k\odot \mathbf x_i \in \mathbb C^{M}$, where $\bar{(\cdot)}$ is an elementwise conjugate, so that $\hat{\mathbf h}_{k}^{*}(\mathbf w)\mathbf f_i=b_{k,i}+\mathbf d_{k,i}^{\mathrm T}\mathbf w$. Thereafter, to express MSE in lifted form, define the augmented matrix $\tilde{\mathbf W}\triangleq\begin{bmatrix}
\mathbf W & \mathbf w\\
\mathbf w^{*} & 1
\end{bmatrix}~(\tilde{\mathbf W}\succeq \mathbf 0,~\mathrm{rank}(\tilde{\mathbf W})=1)$ with $\mathbf W = [\tilde{\mathbf W}]_{1:M,1:M}$. For $\Big|1-u_k^{*}\hat{\mathbf h}_{k}^{*}(\mathbf w)\mathbf f_k\Big|^2=\big|1-u_k^{*}(b_{k,k}+\mathbf d_{k,k}^{\mathrm T}\mathbf w)\big|^2$, define $b_{k,k}^{(0)}\triangleq 1-u_k^{*}b_{k,k},~\mathbf d_{k,k}^{(0)}\triangleq -u_k^{*}\mathbf d_{k,k}$, and $\mathbf G_{k,k}^{(0)}\triangleq
\begin{bmatrix}
\bar{\mathbf d}_{k,k}^{(0)}\mathbf d_{k,k}^{(0)\mathrm T}
& b_{k,k}^{(0)}\bar{\mathbf d}_{k,k}^{(0)}\\
b_{k,k}^{(0)*}\mathbf d_{k,k}^{(0)\mathrm T}
& |b_{k,k}^{(0)}|^2
\end{bmatrix}$. Then $\big|1-u_k^{*}\hat{\mathbf h}_{k}^{*}(\mathbf w)\mathbf f_k\big|^2=\operatorname{tr}\big(\mathbf G_{k,k}^{(0)}\tilde{\mathbf W}\big)$. For $\sum_{i\neq k}\Big|u_k^{*}\hat{\mathbf h}_{k}^{*}(\mathbf w)\mathbf f_i\Big|^2$, define for $i\neq k$: $b_{k,i}^{(1)}\triangleq u_k^{*}b_{k,i},~\mathbf d_{k,i}^{(1)}\triangleq u_k^{*}\mathbf d_{k,i}$, and $\mathbf G_{k,i}^{(1)}\triangleq
\begin{bmatrix}
\bar{\mathbf d}_{k,i}^{(1)}\mathbf d_{k,i}^{(1)\mathrm T}
& b_{k,i}^{(1)}\bar{\mathbf d}_{k,i}^{(1)}\\
b_{k,i}^{(1)*}\mathbf d_{k,i}^{(1)\mathrm T}
& |b_{k,i}^{(1)}|^2
\end{bmatrix}$, which yields $\big|u_k^{*}\hat{\mathbf h}_{k}^{*}(\mathbf w)\mathbf f_i\big|^2=\operatorname{tr}\big(\mathbf G_{k,i}^{(1)}\tilde{\mathbf W}\big)$. Collecting all $\mathbf w$-dependent terms, the joint update becomes
\begin{equation}
\label{prob:Wp_SDP_full_corrected}
\begin{aligned}
&\min_{\tilde{\mathbf W},\mathbf p}~ \sum_{k=1}^{K}\nu_k\Bigg(\operatorname{tr}\big(\mathbf G_{k,k}^{(0)}\tilde{\mathbf W}\big)+\sum_{i\neq k}\operatorname{tr}\big(\mathbf G_{k,i}^{(1)}\tilde{\mathbf W}\big) \\
&~~~~~~~~~~~+|u_k|^2\sum_{i=1}^{K}\Big(\sigma_{d,k}^2\|\mathbf f_i\|_2^2
+\sigma_{r,k}^2\operatorname{tr}(\tilde{\mathbf Q}_i\mathbf W)\Big)\\
&~~~~~~~~~~~+|u_k|^2\sigma_r^2\operatorname{tr}(\tilde{\mathbf R}_k\mathbf W)
+|u_k|^2\sigma_r^2\sigma_{r,k}^2\operatorname{tr}(\tilde{\mathbf T}\mathbf W)\\
&~~~~~~~~~~~+|u_k|^2\sigma_k^2
\Bigg)\\
\mathrm{s.t.}~
&\sum_{i=1}^{K}\operatorname{tr}(\tilde{\mathbf Q}_i\mathbf W)
+\sigma_r^2\operatorname{tr}(\tilde{\mathbf T}\mathbf W)\le P_{\max},\mathrm{rank}(\tilde{\mathbf W})=1,\\
& [\mathbf W]_{mm}\le g_{\max}^2 p_m,
0\le p_m\le 1 ~( \forall m),~\mathbf 1^{\mathrm T}\mathbf p=M_o,\\
&\tilde{\mathbf W}\succeq \mathbf 0,\mathbf W = [\tilde{\mathbf W}]_{1:M,1:M},~[\tilde{\mathbf W}]_{M+1,M+1}=1.
\end{aligned}
\end{equation}
Herein, dropping $\mathrm{rank}(\tilde{\mathbf W})=1$ yields a convex semidefinite program (SDP) in $(\tilde{\mathbf W},\mathbf p)$. Let $(\tilde{\mathbf W}^{\star},\mathbf p^{\star})$ denote the optimal solution of the relaxed SDP. Recall that $\mathbf p^{\star}\in[0,1]^M$ provides a soft indication of the activation likelihood of each FARIS port and is exploited to guide the recovery of an $M_o$-sparse solution $\mathbf w^\star$. Thereafter, $\mathbf w^{\star}$ is obtained via Gaussian randomization and reconstruction procedure~\cite{sdr}.
\subsubsection{Complete AO Framework}
Based on the above developments, we adopt an AO framework to solve~\eqref{prob:WMMSE_relax}. Specifically, starting from an initial feasible point $(\mathbf F^{(0)},\mathbf w^{(0)},\mathbf p^{(0)})$, the AO procedure iteratively performs: i) Update $(\boldsymbol{\nu},\mathbf u)$ via~\eqref{eq:u_nu_update_new}, ii) Update $\mathbf F$ by solving~\eqref{prob:F_sub_new}, and iii) Update $(\mathbf w,\mathbf p)$ by solving~\eqref{prob:Wp_SDP_full_corrected} with dropping $\mathrm{rank}(\tilde{\mathbf W})=1$, and perform Gaussian Randomization to recover an $M_o$-sparse $\mathbf w$. At each AO iteration, the objective is non-decreasing, since each subproblem is solved for the corresponding variable block. Moreover, the surrogate objective is upper-bounded due to the BS and FARIS power constraints. Therefore, the proposed AO algorithm converges to a stationary point of~\eqref{prob:WMMSE_relax}. The complete procedure is summarized in Algorithm~\ref{alg:faris_relax_ao_simple}.
\begin{algorithm}[t]
\caption{AO Framework for Robust FARIS Design}
\label{alg:faris_relax_ao_simple}
\begin{algorithmic}[1]
\Require
$\{\hat{\mathbf h}_{d,k},\hat{\mathbf h}_{r,k},\sigma_{d,k}^2,\sigma_{r,k}^2\}_{k=1}^{K}$,
$\mathbf H_{\mathrm{BR}},\mathbf J,\sigma_r^2,\{\sigma_k^2\}_{k=1}^{K}$,
$P_B,P_{\max},g_{\max},M_o,\epsilon,R$.
\State Initialize $(\mathbf F^{(0)},\mathbf w^{(0)},\mathbf p^{(0)})$ and set $t=0$.
\Repeat
\State Update $\{u_k^{(t+1)},\nu_k^{(t+1)}\}$ by \eqref{eq:u_nu_update_new}.
\State Update $\mathbf F^{(t+1)}$ by solving \eqref{prob:F_sub_new}.
\State Solve~\eqref{prob:Wp_SDP_full_corrected} (drop $\mathrm{rank}(\tilde{\mathbf W})=1$) to obtain $(\tilde{\mathbf W}^\star,\mathbf p^\star)$.
\State Reconstruct $\mathbf w^{(t+1)}$ by Gaussian Randomization.
\State $\mathbf p^{(t+1)} \leftarrow \mathbf p^\star$, $t\leftarrow t+1$.
\Until{the objective converges}
\State \textbf{Output:} $(\mathbf F^{\star},\mathbf w^{\star},\mathbf p^\star) \leftarrow (\mathbf F^{(t)},\mathbf w^{(t)},\mathbf p^{(t)})$.
\end{algorithmic}
\end{algorithm}
\subsection{Computational Complexity}
To analyze the complexity of the proposed AO framework, we compute the complexities of each AO updates. Specifically, the $(\mathbf u,\boldsymbol\nu)$-update requires computing~\eqref{eq:u_nu_update_new} for all $k$ and $i$, resulting in a complexity of $\mathcal O\big(MNK^2\big)$. The $\mathbf F$-update solves the convex QCQP in~\eqref{prob:F_sub_new} with variable dimension $NK$, whose worst-case complexity using a generic interior-point method scales as $\mathcal O\big(I_{\rm qc}(NK)^3\big)$ with $I_{\rm qc}$ iterations~\cite{boyd}. The $(\mathbf w,\mathbf p)$-update consists of forming the SDP coefficients and solving~\eqref{prob:Wp_SDP_full_corrected}. Constructing the required matrices incurs a cost of $\mathcal O(KMN+K^2M^2)$, while solving~\eqref{prob:Wp_SDP_full_corrected} over an $(M+1)\times(M+1)$ semidefinite matrix variable has worst-case complexity $\mathcal O(I_{\rm sdp}M^{4.5})$ with $I_{\rm sdp}$ iterations~\cite{sdr}, followed by Gaussian randomization with cost $\mathcal O(RM^2)$ for $R$ random trials. 
Consequently, the total worst-case computational complexity of the proposed AO algorithm over $T_{\rm AO}$ iterations is given by $\mathcal O\left(T_{\rm AO}\Big[I_{\rm qc}(NK)^3 + (K^2+R)M^2 + I_{\rm sdp}M^{4.5} + MNK^2\Big]\right)$.

\section{Simulation Results}
Unless otherwise stated, we consider a scenario in which BS equipped with $N=16$ antennas serves $K=4$ users. The FARIS is configured as $M=64$ candidate elements, among which $M_o=16$ elements are activated, each with $g_{\max}=40$~dB~\cite{aris5}. The BS and the FARIS are located at the origin and at $(250,20)$~m, respectively, while the users are uniformly distributed within a circular region centered at $(500,0)$~m with a radius of $10$~m. The noise powers are set to $\sigma_r^2=\sigma_k^2=-90$~dBm, and the BS transmit and the maximum FARIS power are given by $P_B=P_{\max,t}=25$~dBm with $(P_{\mathrm{DC}}, P_c)=(-5 -10)$~dBm. We assume Rician fading for the estimated channels, and the variances of the CSI errors are uniformly set to $\sigma_{j,k}^2 = \frac{\delta^2}{N}\mathbb E[\|\Delta \mathbf h_{j,k}\|_2^2]$ with $\delta=0.2$. To assess the performance of the proposed robust design, we employ the sum-rate in~\cite[(16)]{aris8} as the primary performance metric. For performance comparison, we consider the robust ARIS design in~\cite{aris8}, FRIS/RIS-aided design with AO framework, and the scheme without reconfigurable surfaces.

\begin{figure}[t]
    \centering
    \subfloat[]{%
        \includegraphics[width=0.2\textwidth]{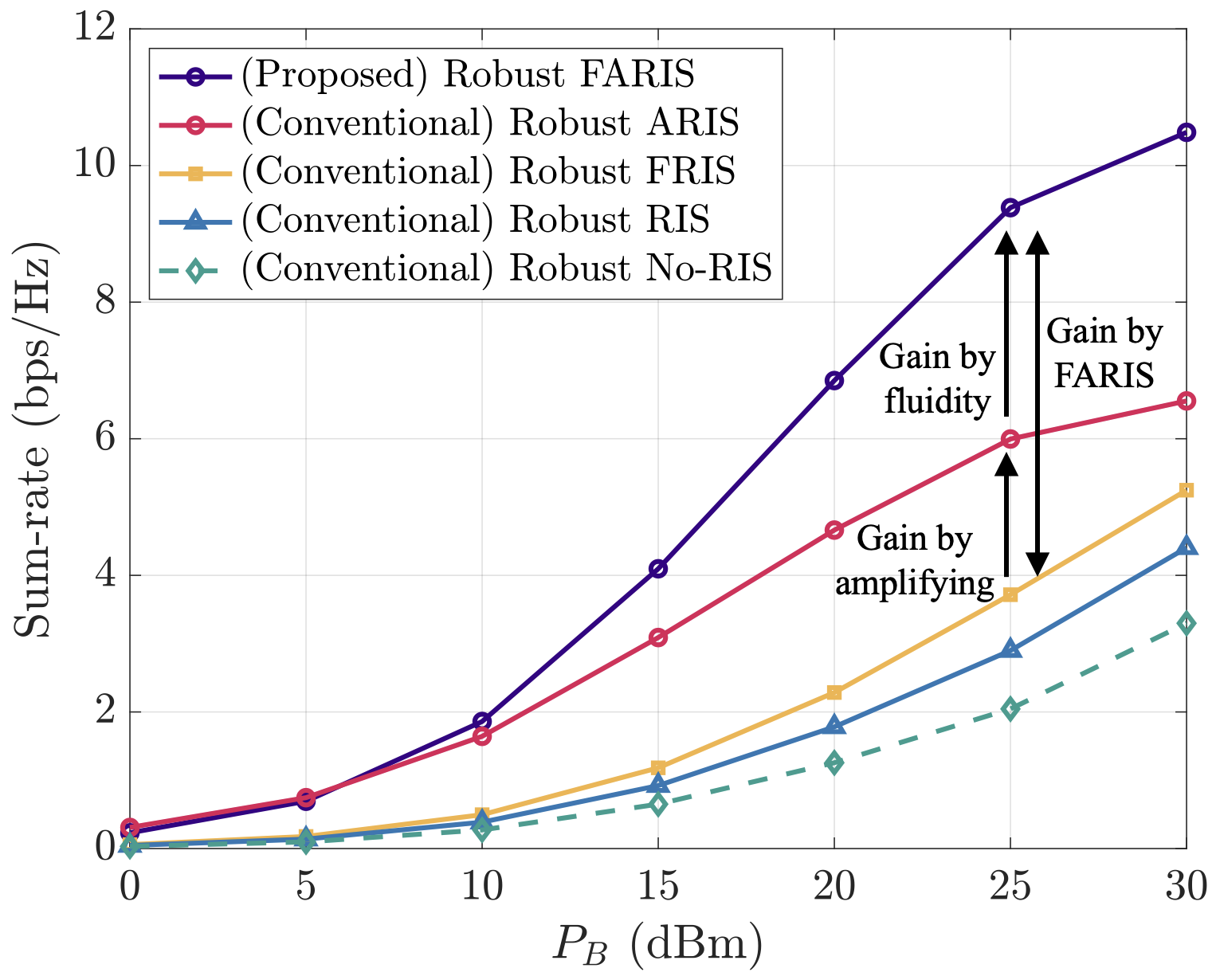}
        \label{fig_pb}%
    }
    \subfloat[]{%
        \includegraphics[width=0.2\textwidth]{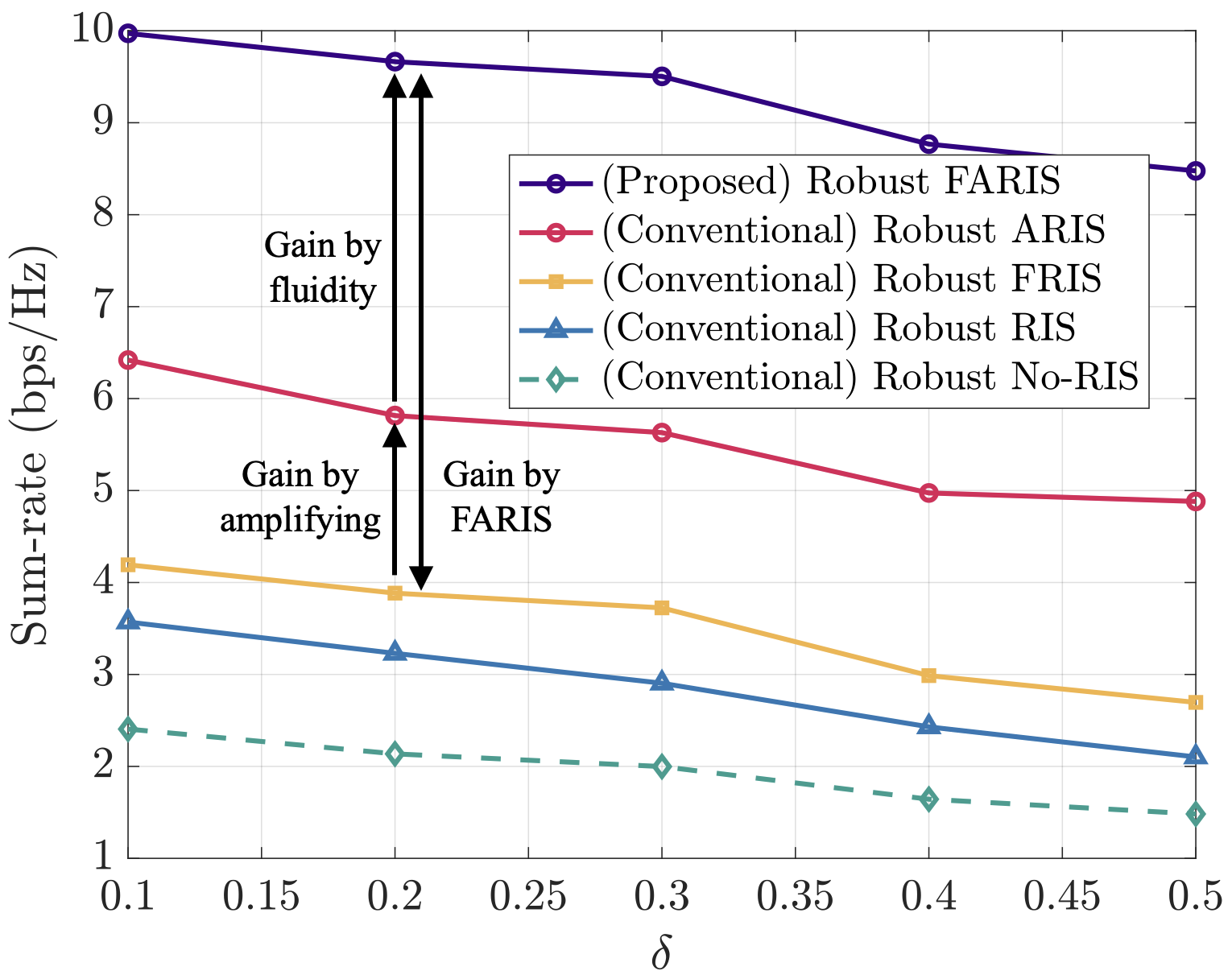}
        \label{fig_err}%
    }
    \caption{Sum-rate according to (a) $P_B$ and (b) $\delta$.}
    \label{fig_1}
\end{figure}
Fig.~\ref{fig_pb} depicts the sum-rate performance as a function of $P_B$. Notably, the proposed robust FARIS exhibits a significantly steeper performance improvement than the conventional ARIS, FRIS, RIS, and No-RIS benchmarks. Even in the transmit-power-limited regime imposed by the FARIS radiated power constraint for FARIS/ARIS~\cite{aris8}, the proposed FARIS consistently outperforms the benchmark schemes. This result demonstrates that the proposed FARIS architecture can effectively leverage increased transmit power through the joint optimization of beamforming and active element selection, even in the presence of CSI uncertainty.

Fig.~\ref{fig_err} illustrates the average sum-rate performance as a function of $\delta$. As $\delta$ increases, the variances of the channel estimation errors become larger, which amplifies the mismatch between the true and estimated channels. Consequently, the sum-rate performance of all considered schemes degrades monotonically with increasing $\delta$. Nevertheless, the proposed robust FARIS consistently outperforms the conventional ARIS, FRIS, and RIS schemes, which highlights the effectiveness of the proposed robust design in mitigating the adverse impact of CSI uncertainty through the joint exploitation of active amplification and robustness-aware optimization.

\begin{figure}[t]
    \centering
    \subfloat[]{%
        \includegraphics[width=0.2\textwidth]{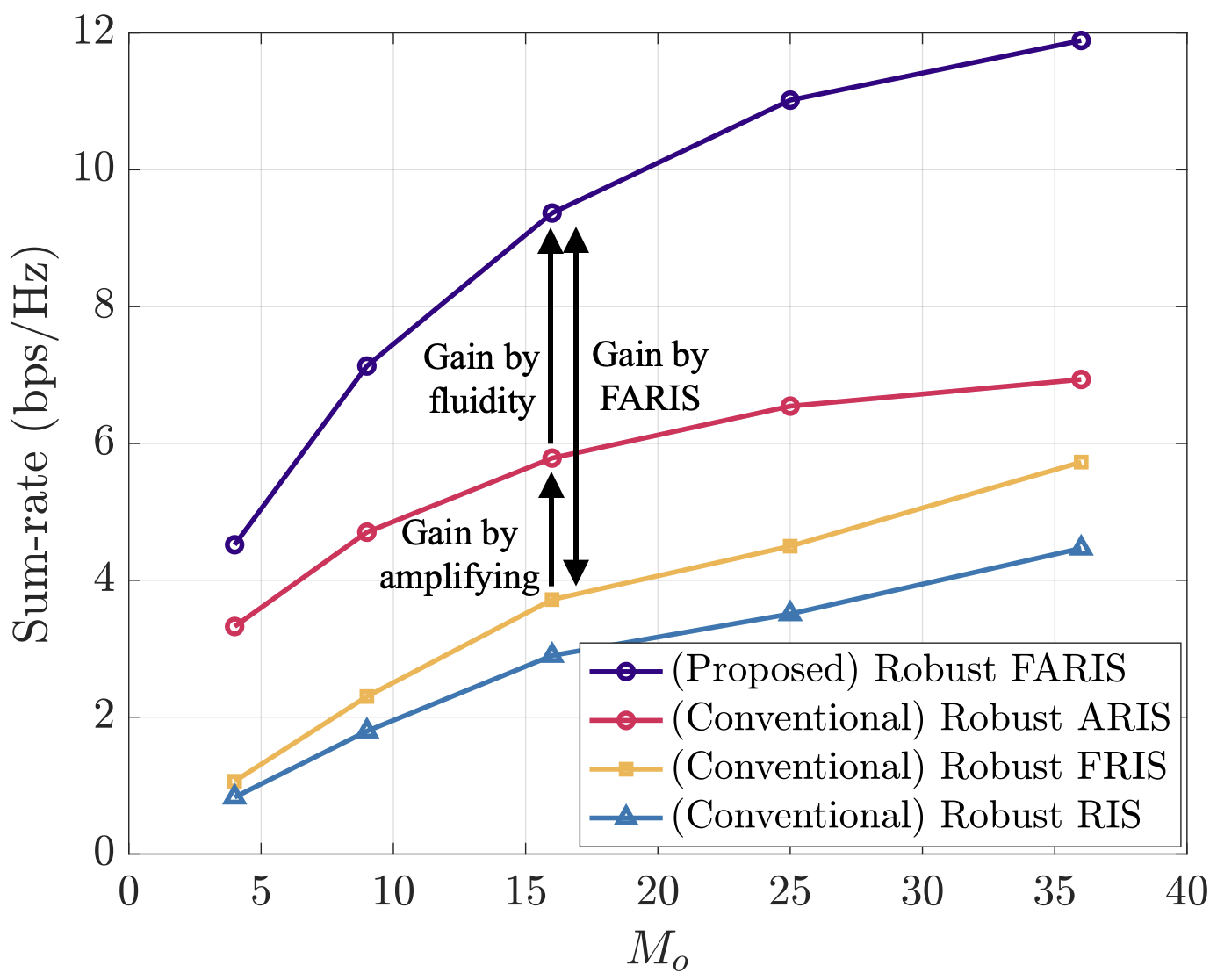}
        \label{fig_mo}%
    }
    \subfloat[]{%
        \includegraphics[width=0.2\textwidth]{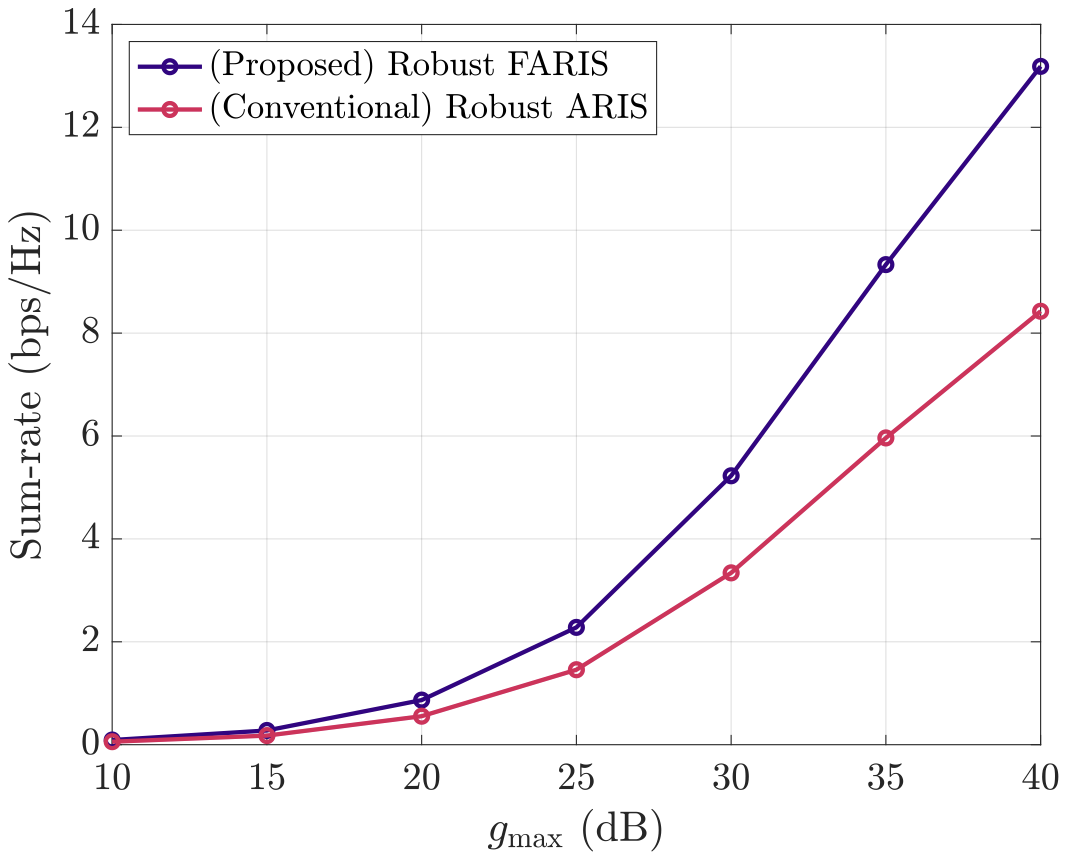}
        \label{fig_gain}%
    }
    \caption{Sum-rate according to (a) $M_o$ and (b) $g_{\max}$.}
    \label{fig_2}
\end{figure}
Fig.~\ref{fig_mo} shows the average sum-rate performance with respect to~$M_o$. Herein, we assumed $M_o=M$ for ARIS and RIS. As $M_o$ increases, the proposed robust FARIS achieves a significant sum-rate improvement due to the enhanced spatial DoF and array and amplification gain by efficiently utilizing a larger $M_o$~\cite{FARIS}. In contrast, the sum-rate of the conventional robust ARIS and RIS exhibits less improvement compared to FARIS, indicating its limited capability to exploit fluidity or both fluidity and amplification, respectively, over elements under practical constraints. The FRIS benchmark also shows slower growth, since passive reflection without amplification cannot fully leverage the increased $M_o$.

Fig.~\ref{fig_gain} depicts the average sum-rate as a function of~$g_{\max}$. It is observed that the sum-rate achieved by the proposed robust FARIS increases rapidly with~$g_{\max}$, and in particular, the performance gap between the proposed FARIS and the conventional robust ARIS widens. This is because the proposed FARIS inherently offers additional spatial DoF through active element selection~\cite{FARIS}, whose benefits are further strengthened as $g_{\max}$ increases, thereby leading to a more pronounced performance gain compared to the ARIS architecture.
\section{Conclusion}
This paper showed that FARIS fundamentally enlarges the robustness and performance design space of multi-user downlink systems by coupling {active amplification} with {fluid port utilization}, so that DoF creation and link-strength enhancement can be jointly exploited even under statistical CSI uncertainty. A key lesson was that robust FARIS design is not a simple add-on to ARIS/FRIS beamforming: the amplification-induced noise and radiated-power coupling, together with the sparsity-driven port utilization structure, must be co-designed with the BS beamformer to avoid robustness loss and to preserve multi-user interference control. The proposed WMMSE-based reformulation and AO decomposition clarified a principled pathway for such co-design, enabling stable updates of the beamformer and surface variables while explicitly incorporating CSI-error contributions from both direct and FARIS-assisted links. These findings indicate that FARIS constitutes a promising architecture for 6G, especially in scenarios requiring both robustness and high spectral efficiency.
\bibliographystyle{IEEEtran}
\bibliography{IEEEexample}

\end{document}